\documentstyle[aps]{revtex}
\begin{document}

\draft

\title{Thick domain wall spacetimes with and without reflection symmetry}

\author{Alejandra Melfo$^{(1,2)}$
, Nelson Pantoja$^{(1)}$ 
  and  Aureliano Skirzewski$^{(1,2)}$ 
\footnote{e-mail: melfo@ula.ve, pantoja@ula.ve, skirz@ula.ve}
}

\address{$^{(1)}$ {\it Centro de Astrof\'{\i}sica Te\'orica, Universidad de
Los Andes, M\'erida, Venezuela }}

\address{$^{(2)}${\it International Centre for Theoretical Physics,
34100 Trieste, Italy }}

\maketitle

\begin{abstract}

We show that the spacetimes of domain wall solutions to the
coupled Einstein-scalar field equations with a given scalar field
potential fall into two classes, depending on whether or not
reflection symmetry on the wall is imposed. Solutions with
reflection symmetry are dynamic, while the asymmetric ones are
static. Asymmetric walls are asymptotically flat on one side and
reduce to the Taub spacetime  on the other. Examples of asymmetric
thick walls in D-dimensional spacetimes are given, and results on
the thin-wall limit of the dynamic, symmetric walls are extended
to the asymmetric case. The particular case of symmetric, static
spacetimes is considered and a new family of solutions, including
previously known BPS walls, is presented.

\end{abstract}


\section{Introduction}

 Scalar fields as sources for the Einstein field equations have been subject
of recent interest. Configurations with a plane-parallel symmetry,
i.e. scalar field walls, are particularly appealing, since they can be
 topologically
stable. Although in four dimensions they are in conflict with
standard cosmology \cite{k76}, in theories with extra dimensions
the possibility of realizing four-dimensional gravity in a
$3$-brane or thin domain wall \cite{rs99}, already suggested in
\cite{rs83}, has attracted wide attention.

Self gravitating domain walls are solutions to the coupled
Einstein-scalar field system, with a potential $V(\phi)$
possessing a (spontaneously broken) discrete symmetry. The coupled
system of equations is solved for the field $\phi$ and the  metric
tensor components $g_{ab}$. A topological charge for the wall is
obtained by mapping the values of $\phi$ at spatial infinity to
the vacuum manifold in a non-trivial way. To simplify matters, one
looks for solutions representing static domain walls, that is, one
requires the energy and momentum density to be time-independent.
Because the field is required to take values at the (degenerate)
minima of the potential at spatial infinity on both sides of the
wall, the energy and momentum densities are invariant under
reflections on the wall's plane.

 These features of the domain wall solution, staticity and
reflection symmetry, need not be shared by the spacetime metric.
As shown by Vilenkin \cite{v83}, the most general static vacuum
solution with plane-parallel and reflection symmetry obtained by
Taub \cite{t51} cannot be the external spacetime of a true domain
wall source (in the absence of a cosmological constant). The first
vacuum solution for a spacetime containing an infinitely thin
sheet of scalar field compatible with the equation of state of a
domain wall was obtained in \cite{v83}. The spacetime is not
static, but has a de Sitter expansion on the wall's plane. Since
then, solutions to the coupled Einstein-scalar field equations
with a symmetry breaking potential have been found
\cite{g90,gm99,cgs93}. In the absence of a cosmological constant,
those solutions reduce, in the thin-wall limit, to the dynamic
solution of \cite{v83}. On the other hand, one can have a static
metric as long as the wall is allowed to interpolate between
Anti-de-Sitter (AdS) asymptotic vacua, the so-called BPS walls
\cite{st99,dfgk00}.

Asymmetric thin domain wall spacetimes (or asymmetric brane-world
scenarios) in which the reflection symmetry along the extra
dimension was broken by gluing two AdS spacetimes with different
cosmological constants have been considered in
\cite{k99,i00,dd00,stw00}, and their possible realization by the
introduction of a gauge form field has been proposed in \cite{bc01}.
For thick domain wall spacetimes, a less known result is that the
condition of staticity of the metric can be maintained even in the
absence of a cosmological constant term, if one is willing to
sacrifice reflection symmetry. We only know of an example in the
literature so far, for a domain wall in four dimensions, that
interpolates between a Minkowski and a Taub spacetime \cite{gm99}.
This solution to the coupled Einstein-scalar field equations has
the peculiarity of having the same scalar field potential
$V(\phi)$ as the well-known dynamic solution of Ref. \cite{g90},
which has prompted us to study this issue in more detail, and to
consider the extension to higher-dimensional spacetimes.

In what follows we show that domain wall solutions to the
Einstein-scalar field equations in D dimensions, for a scalar
field with a given potential $V(\phi)$ fall into two classes: (a)
dynamic, symmetric solutions, with a de Sitter expansion on the
wall's plane, and (b) static, asymmetric solutions, interpolating
between  Minkowski and Taub spacetimes (rather, their
D-dimensional equivalents). Static and symmetric solutions are
obtained as particular cases, and they represent walls embedded in
a spacetime with a cosmological constant.

These two classes of spacetimes are solutions to the equations with the
same potential $V(\phi)$ and the same wall profile $\phi(\xi)$ (where
$\xi$ is the bulk coordinate). They are found with the
same boundary conditions on $\phi$ at infinity, and their energy
density is in both cases static and reflection symmetric. Thus, the
metric is not uniquely determined, but depends on  subsidiary
conditions imposed on its components.

Both classes of spacetimes cannot be related with a coordinate
transformation, but there is a one-to-one correspondence between
the dynamic and the asymmetric solutions. We take advantage of
this by applying a recently reported method \cite{gmp02} for
solving the coupled Einstein-scalar field system to obtain
asymmetric solutions, by appropriately scaling the vacuum
solutions. Results reported in \cite{gmp02} concerning the
thin-wall limit of these solutions are shown to be valid in the
asymmetric case. The results are then extended to the general case
of a D-2 dimensional brane in a D-dimensional spacetime.

In order to obtain further examples, we show how new solutions can
be obtained by a different way of scaling vacuum solutions. A
particularly interesting class of static solutions representing a
parametric family of ``double'' walls, i.e. walls with energy
density concentrated in two parallel sheets  is considered in some
detail. These walls reduce to a known BPS thick domain wall
\cite{g99} for a particular value of the parameter. Other
solutions, for theories with  less appealing scalar field
potentials, are also presented.

\section{Dynamic vs. asymmetric solutions}

The most general metric for a 5-dimensional spacetime with a
plane-parallel symmetry can be written as \begin{equation} g_{ab}=
e^{2\mu(\xi)}[-dt_a dt_b + C(\xi,t)^2 dx^i_a dx^i_b ] +
e^{2\nu(\xi)}d\xi_a d\xi_b \label{genmetric} \end{equation} where latin
indices run over the spatial variables on the brane. The function
$\mu(\xi)$ in (\ref{genmetric}) is redundant, since only two
functions are needed in general. We will keep it, however, and
choose it conveniently as a function of $\nu(\xi)$ and $C(\xi,t)$
later.
We look for solutions to 
\begin{equation} G_{ab} + g_{ab}\Lambda = T_{ab}\; \; ; \quad \;\;\;\;
  T_{ab} =
\partial_a\phi\partial_b\phi -
g_{ab}(\frac{1}{2}\partial^c\phi\partial_c\phi + V(\phi)),
\label{einstein} \end{equation} satisfying the requirements
\begin{enumerate}
\item $\phi = \phi(\xi)$,
\item $V(\phi)$ has a (spontaneously broken) discrete symmetry
\item $\phi(\xi)$ takes different values at two different minima of $V(\phi)$
  for $|\xi|\to\infty$
\item $\phi(\xi)'^2$ is symmetric under reflections in the $\xi=0$
  plane\footnote{
Here and in what follows  primes denote derivative
  respect to $\xi$ and dots derivatives respect to $t$.}.
\end{enumerate}
Following the usual strategy, we will first find $C(\xi,t)$ by imposing
 the requirements of staticity and reflection symmetry of $\phi(\xi)'^2$
 and $V(\phi(\xi))$, given by
\begin{eqnarray} \phi(\xi)'^2 &=& e^{2\nu} (G^4_4 - G^0_0),\\
 V(\phi(\xi)) + \Lambda &=&
-\frac{1}{2} (G^4_4 + G^0_0)  , \end{eqnarray} and then look for solutions
$\{\phi(\xi), V(\phi + \Lambda\}$
 in terms of the ``warp factor'' $\nu(\xi)$.
We have from (\ref{einstein})
\begin{equation}
G^0_4 = 3 e^{-2\mu}\frac{\dot C'}{C} =0 ,
\end{equation}
therefore $C(\xi,t) $ is the sum of a function for $t$ and a function
 of $\xi$.
With this, by requiring
\begin{eqnarray}
G^0_0 - G^1_1 = -2 e^{-2\mu}\left( \frac{\dot C}{C}\right)^. +
e^{-2\nu}\left[\left(\frac{C'}{C}\right)' + 
  \frac{C'}{C}(4\mu' - \nu' + 3 \frac{C'}{C})\right] =0
\label{cero}
\end{eqnarray}
for arbitrary $\nu(\xi)$, two types of solutions are possible:
\begin{description}
\item[{\bf A}] Static solutions, with  $C= C(\xi) \equiv e^{g(\xi)}$ .

Since the most general static metric can be written in terms of
two functions,  we can conveniently set
\begin{equation}
  \mu = \frac{1}{4}\nu - \frac{3}{4} g.
 \end{equation}
In this case (\ref{cero}) is integrated to give $g(\xi) =
\beta\xi$ and  we have \begin{eqnarray} \phi_{\bf A}'^2 &=& \frac{3}{4}[\nu'^2
-\beta^2 - \nu'']\\ 
V(\phi)_{\bf A} &=& -\frac{3}{8} e^{-2\nu}
\nu'' -  \Lambda , \end{eqnarray}
 static as required. They will also be
reflection symmetric if so is  $\nu(\xi)$.

\item[{\bf B}]  Dynamic solutions, with $C= C(t) \equiv e^{h(t)}$

Eq. (\ref{cero}) gives $h(t)=\beta t$, and
we can now  set $\mu = \nu $,
obtaining
\begin{eqnarray}
\phi_{\bf B}'^2 &=& 3[\nu'^2 -\beta^2 - \nu''] \\
V(\phi)_{\bf B} &=&- \frac{3}{2} e^{-2\nu} [\nu'' + 3\nu'^2 - 3
  \beta^2] -  \Lambda  .
\end{eqnarray}\end{description}

While we have ensured that the field's gradient and potential are
static and symmetric under reflections in the $\xi=0$ plane, the
spacetimes of solutions {\bf A} and {\bf B} are not. The metric of
solutions {\bf A} is manifestly asymmetric, although static \begin{equation}
{(g_{\bf A})_{ab}}= e^{\nu(\xi)/2 - 3\beta \xi/2}[-dt_a dt_b +
e^{2\beta\xi} dx^i_a dx^i_b ] + e^{2\nu(\xi)}d\xi_a d\xi_b
\label{metricA} \end{equation} Instead, the metric in solutions {\bf B} is
dynamic, but symmetric \begin{equation} {(g_{\bf B})_{ab}}=
e^{2\nu(\xi)}[-dt_a dt_b + e^{2\beta t}dx^i_a dx^i_b  ] +
e^{2\nu(\xi)}d\xi_a d\xi_b \label{metricB} \end{equation} Solutions of type
{\bf B} are encountered in the literature, both in 4 and 5
dimensional spacetimes \cite{g90,g99,kks02,gmp02},  while only one
example of those of  type {\bf A} has been  discussed, in 4
dimensions \cite{gm99}.

The coupled system of equations (\ref{einstein}) has
now to be solved by proposing a warp factor such that $\{\phi(\xi),
V(\phi) \}$ can be integrated. The remarkable point is that the
equation for $\phi$ is the same in both cases. Therefore the warp
factors for vacuum solutions in both spacetimes, obtained by
integrating the equations $\phi'^2=0$ for $\nu(\xi)$, are the
same. However the spacetimes will have different cosmological
constants.

Now, in Ref. \cite{gmp02} we presented a method for generating
solutions to the system (\ref{einstein}) (in 4
dimensions) with a spacetime of type {\bf B} by scaling the
vacuum solutions. Specifically, we showed that if $\nu_0(\xi)$ is
a vacuum solution with a (non-null) cosmological constant
$\Lambda_0$, the system can be integrated with the function
 \begin{equation}
\nu(\xi) = \delta \nu_0(\xi/\delta) \end{equation}
where $0<\delta<1$ This holds true for  a
higher-dimensional wall, and, more importantly, for spacetimes of
type {\bf A}. We obtain 
\begin{equation} \phi=\sqrt{\frac{2 \Lambda_0}{a}}
\sqrt{\delta(1-\delta)}
\int_{\xi_0}^{\xi/\delta}{e^{\nu_0(\omega)}
d\omega} \end{equation} 
\begin{equation} V(\phi) = \frac{[1+\delta(a-1)]}{a}
\frac{\Lambda_0}{\delta}\exp[2\nu_0(\xi/\delta)(1-\delta)] \; ;
\quad  \Lambda = 0 \end{equation} where  $a=1$ for case {\bf A} and $a=4 $
for case {\bf B}.

So, it is not only possible to generate solutions for asymmetric
spacetimes by using this method: the  point is that the scalar
field and the potential in the asymmetric and the dynamic cases
differ by an overall constant only. Therefore, given a theory with
a scalar potential, two solutions can be found to the
Einstein-scalar field equations with essentially the same scalar
field configuration, but representing different spacetimes.

To further illustrate this point, consider the solution found by
scaling the vacuum solution $\nu_0(\xi) = -\ln[\cosh(\beta\xi)]$,
$\Lambda_0=3 a^2 \beta^2/8$: 
\begin{equation} \nu(\xi) = -\delta
\ln[\cosh(\beta\xi/\delta)] \label{nucosh} \end{equation} We have
\begin{eqnarray} \phi(\xi) &=&  \phi_0
\tan^{-1}[\sinh(\beta\xi/\delta)]  , \quad 
\phi_0
=\sqrt{\frac{2\Lambda_0}{a}}\frac{ \sqrt{\delta(1-\delta)}}{\beta};
\label{phicosh} \\
 V(\phi) &=& \frac{ [1+\delta(a-1)]}{\delta}
\frac{\Lambda_0}{a} [\cos(\phi/\phi_0)]^{2(1-\delta)}. 
\label{vecosh} \end{eqnarray}
With the dynamic metric of case {\bf B}, this is just the
5-dimensional analogue of Goetz's solution \cite{g90,kks02}. With
the asymmetric metric of case {\bf A}, this is the 5-dimensional
analogue of the solution found in \cite{gm99}.

We now wish to make contact with the brane-world scenarios and
take the thin wall limit of (\ref{metricA},\ref{nucosh}) and its
curvature tensor fields. In Ref.\cite{gmp02}, it was  shown that
the domain wall spacetime with metric given by
(\ref{metricB},\ref{nucosh})  has a well-defined thin wall limit. The
corresponding asymmetric wall shares this property.

It is easy to see that (\ref{metricA},\ref{nucosh}) is a regular
metric in the sense of \cite{gt87}. We have that both $g_{ab}$ and
$(g^{-1})^{ab}$ are locally bounded. Further, with $\eta_{ab}$ the
ordinary Minkowski metric in 5 dimensions, we find that the weak
derivative in $\eta_{ab}$ of $g_{ab}$ exists and is locally square
integrable. Hence $g_{ab}$ can be considered as a distributional
metric and its curvature tensor fields make sense as tensor
distributions. Taking the $\delta \to 0$ limit (in the sense of
distributions) we find
 \begin{eqnarray} \lim_{\delta\to 0}{{g_{ab}}} &=& e^{-
\beta (|\xi| + 3 \xi)/2}[-dt_a dt_b + e^{2\beta\xi}dx^i_a dx^i_b ]
+ e^{-2\beta|\xi|}d\xi_a d\xi_b \label{thinwall} \\
\lim_{\delta \to 0} G^a_{\,\,b} &=& -\frac{3}{2} \beta
\delta(\xi)[\partial^a_t dt_b +  \partial_{x^i}^a dx^i_a] \end{eqnarray}
 For
$\xi>0$, (\ref{thinwall}) is just the Minkowski spacetime, while
for $\xi<0$ it is the 5-dimensional analogue of the Taub solution
\cite{t51}. By performing two different coordinate
transformations,  the metrics on both sides of the wall can be
cast in a more familiar form (see \cite{gm99} for this, and for a
detailed analysis of geodesics in  the 4-dimensional case). Hence,
the spacetime with $g_{ab}$ given by (\ref{metricA},\ref{nucosh})
is an explicit realization of an asymmetric thick domain wall
spacetime with a well defined thin domain wall limit.

\section{Extension to $D$ dimensions}
It is straightforward to extend these results for a thick
($D$-2)-brane embedded in a D-dimensional spacetime. Writing the
metrics as 
\begin{eqnarray} {(g_{\bf A})_{ab}}&=& e^{2(\nu(\xi) -(D-2)
\beta\xi)/(D-1)}[-dt_a dt_b + e^{2\beta\xi}dx^i_a dx^i_b ]
 + e^{2\nu(\xi)}d\xi_a d\xi_b \label{metricAD}\\
{(g_{\bf B})_{ab}}&=& e^{2\nu(\xi)}[-dt_a dt_b + e^{2\beta
t}dx^i_a dx^i_b + d\xi_a d\xi_b] \label{metricBD} \end{eqnarray} we get 
\begin{eqnarray}
\phi(\xi)'^2 &=& a_D \frac{(D-2)}{(D-1)}[\nu'^2 -\beta^2 - \nu''] \\
V(\xi) &=& - \frac{a_D}{2}\frac{(D-2)}{(D-1)} e^{-2\nu} [\nu''
+(a_D-1)( \nu'^2
-  \beta^2)] -  \Lambda  .
\end{eqnarray}
where now $a_D=1$ for case {\bf A} and $a_D=D-1$ for case {\bf B}.
In particular, with $\nu(\xi)$ given by (\ref{nucosh}),
solutions (\ref{phicosh}) and (\ref{vecosh}) for the field and the 
potential are found,
 with $\Lambda_0=\beta^2 a_D^2(D-2)/2(D-1)$ .

One can now proceed to obtain other solutions by scaling all the
vacuum solutions, namely
 \begin{eqnarray}
\nu = \ln[\cosh(\beta\xi)]\;\; &;& \;\;
 \Lambda_0 = a_D^2 \frac{\beta^2}{2}\frac{D-2}{D-1} \label{coshvacio}\\=
\pm \beta\xi\;\; &;&\;\; \Lambda_0 = 0  \label{expvacio}\\=
\ln[\sinh(\beta\xi)] \;\;&;&\;\; \Lambda_0 = a_D^2 \frac{\beta^2}{2}\frac{D-2}{D-1}
\end{eqnarray}
when $\beta\neq 0$, and
\begin{eqnarray}
\nu= \ln(\alpha \xi)\;\; &;&\;\; \Lambda_0 = a_D \frac{\alpha^2}{4} \label{lnvacio}\\ =0\;\; &;&\;\; \Lambda_0
=0 \label{nullvacio}
\end{eqnarray}
when $\beta=0$, where $\alpha$ is an integration constant. Notice that
with metric
{\bf A}, the two solutions  (\ref{expvacio}) correspond to the
D-dimensional analogues of
Minkowski and Taub spacetimes respectively.
 A number of solutions were found in Ref. \cite{gmp02} for the
dynamic spacetime of case {\bf B} by scaling these vacuum solutions. For each
 of these solutions there is a  corresponding asymmetric one. However, it
 was shown that among them  the only domain wall solution, meaning one that
interpolates between two minima of the potential, is
(\ref{nucosh}-\ref{vecosh})  and this is also true for the asymmetric
solutions.

It should be stressed that the asymmetric thick branes considered arise
as solutions to the Einstein-scalar field equations with a {\em
symmetric potential} possessing a $Z_2$ symmetry. Furthermore, the
spacetime asymmetry cannot be eliminated by a coordinate change.
This can be readily seen from the Kretschmann scalar
($K=R_{\alpha\beta\gamma\delta}R^{\alpha\beta\gamma\delta}$), \begin{eqnarray}
K_{\bf A}&=& \frac{2 e^{-4\nu}}{(D-1)^3}[2 (D-1)^2\nu''^2 -
4(D-1)(D-2)\nu''(\nu'^2-\beta^2) +(D-2)(2 D - 3)(\nu'^2 -
\beta^2)^2 \nonumber \\& & + 2\beta^2(D-2)^2(D-3)(\nu' + \beta)^2]
\end{eqnarray} which for the solution (\ref{nucosh}-\ref{vecosh})is
 manifestly asymmetric
\begin{eqnarray}
K_{\bf A} &=& \frac{2}{(D-1)^3} \beta^4 [\cosh(\beta\xi/\delta)]^{-4(1-\delta)}\left[-2 \frac{1}{\delta^2}(D-1)^2
- 4\frac{1}{\delta}(D-1)(D-2)+(D-2)(2 D - 3)\right. \nonumber \\
& &\left. +2(D-2)^2(D-3) \cosh^2(\beta\xi/\delta) e^{-2\beta\xi/\delta} \right]
\end{eqnarray}
and diverges  as $\xi\to\infty$, but goes to zero for
$\xi\to +\infty$.
 The asymmetry is not present in the Ricci scalar, which vanishes for $|\xi|\to\infty$.
 The corresponding solutions of type ${\bf B}$, on the other hand, are asymptotically flat.
Notice that while the asymmetric solutions are static, they are not
in general BPS domain walls.

The fact that the scalar potential for these two different solutions
is the same is a consequence of the scaling procedure we have
followed.  In the next section, we  generate other
thick domain wall  solutions that do not share this property, by proposing
a different type of scaling.

\section{A symmetric, static family of walls}

The warp factor for a thick domain-wall solution can be obtained
{\em via} scaling of the vacuum solutions warp factors in more
than one way. A very useful one is the following: take two
different vacuum solutions with warp factors $\exp(\nu_1)$ and
$\exp(\nu_2)$ respectively, and define the warp factor for the
thick wall as \begin{equation} \nu(\xi) = - \frac{1}{2s} \ln\left[ \exp(-2s
\nu_1) + \exp(-2s \nu_2) \right] \label{nuscale} \end{equation} It turns out
that the Einstein-scalar field equations can  always be integrated
with (\ref{nuscale}.) Naturally, this scaling  will provide
asymmetric as well as dynamic solutions, and as could be expected,
it will work for the spacetimes of arbitrary dimensions considered
in the previous section. In \cite{kks02} this type of scaling has
been used for a pair of vacuum solutions in a 5-dimensional
spacetime of type {\bf B}.

We get solutions with metrics (\ref{metricAD}) or (\ref{metricBD})
for $\phi(\xi)$ and $V(\xi)$ as 
\begin{eqnarray} \phi(\xi)&=& 
\sqrt{\frac{(D-2) a_D}{D-1}} \frac{\sqrt{2s-1}}{2s}
\tan^{-1}[\sinh(s\Delta\nu)] \label{phiscale}\\
 V(\xi)&=&
\frac{a_D}{2}\frac{D-2}{D-1}e^{2\nu}\left\{\frac{\cosh^{-2}(s\Delta\nu)}{4} \left[
(2s-1)(\Delta\nu)'^{2} -  a_D(\nu_1' e^{-s\Delta\nu}
+ \nu_2'e^{s\Delta\nu})^2\right] + a_D\beta^2
 \right\}\label{vscale}
\end{eqnarray}
where  $\Delta\nu \equiv \nu_2 - \nu_1$. By choosing the vacuum
solutions (\ref{expvacio}), the result (\ref{nucosh}-\ref{vecosh}) is
recuperated. In this case, the parameter $s$ plays the role of the
inverse of the wall's thickness, $\delta^{-1}$, but this is not
true in general.

The particular case of symmetric, static solutions is found by
using vacuum solutions with $\beta=0$, namely taking \begin{equation} \nu =
-\frac{1}{2s}\ln( 1 + (\alpha \xi)^{2s} ). \label{nustat} \end{equation} 
We get, for D=5, 
\begin{equation} \phi = \phi_0\tan^{-1}(\alpha^s\xi^s) ,  \quad
\phi_0=\frac{\sqrt{3(2s-1)}}{s} ,
\end{equation} \begin{equation}
V(\phi) + \Lambda =3 \alpha^2\sin(\phi/\phi_0)^{2-2/s}[\frac{2 s + 3}{2}
\cos^2(\phi/\phi_0) - 2] \label{vstat};\end{equation}
so that $\Lambda = - 6 \alpha^2$. In this case, the parameter $s$ 
 cannot be identified with the wall's
inverse thickness. Solutions exist only for  $s$ a positive
integer, and for $s$ even they are not domain walls, since the
field takes values at infinity at the same minimum of the
potential. For $s=1$, this solution has been presented in
\cite{g99} in 5 dimensions. A change of coordinates allows one to
identify it with the  regularized version of the usual
Randall-Sundrum brane. For other (odd) values of $s$, the
potential has a local minimum between two global ones. In a region
around the origin, the field takes values at this local minimum,
and falls to (different) global minima at spatial infinity.

Let us take  a closer look at the solutions with $s$ odd. We would
like to explore the thin-wall limit of these configurations.
Following \cite{gmp02}, we introduce a new parameter $\delta$ by
scaling the solutions (\ref{nustat}) so that the metric is now
\begin{equation} _\delta g_{ab} =\left[ 1 +
\left(\frac{\alpha\xi}{\delta}\right)^{2s}  \right]^{-\delta/s}
(-dt_a dt_b + dx^i_a dx^i_b ) + \left[1 +
\left(\frac{\alpha\xi}{\delta}\right)^{2s}  \right]^{-1/s} d\xi_a
d\xi_b. \label{deltametric} \end{equation}
Notice that the scaling is performed so that this is still a
solution to the Einstein-scalar field equations with
\begin{equation}
\phi(\xi) =\phi_0 \tan^{-1}(\frac{\alpha^s\xi^s}{\delta^s}),
\quad \phi_0= \frac{\sqrt{3\delta (2s
-1)}}{s}, 
\end{equation} 
\begin{equation}
 \quad V(\phi) + \Lambda =3
\alpha^2\sin(\phi/\phi_0)^{2-2/s}[\frac{2 s + 4\delta -1}{2\delta}
\cos^2(\phi/\phi_0) - 2], \end{equation}
and
\begin{eqnarray} G^\xi_{\,\,\xi} &=& 6\alpha^2\left[ 1 +
\left( \frac{\alpha\xi}{\delta} \right)^{-2s} \right]^{1/s -2}
\label{deltag11} \\ G^t_{\,\,t} &=& 6\alpha^2\left[ 1 + \left(
\frac{\alpha\xi}{\delta} \right)^{-2s} \right]^{1/s -2} \left\{ 1
+ \frac{1 - 2s}{2\delta}
\left(\frac{\alpha\xi}{\delta}\right)^{-2s} \right\}
\label{deltag00} \end{eqnarray} 
The function $-G^t_{\,\,t}$, {\em i.e.} the
energy density, has two maxima at \begin{equation} \xi_\pm = \pm \delta\,
[(s-1)/(s+2\delta)]^{1/(2s)} \end{equation} and the wall can in this sense
be considered a ``double wall'' for $s > 1$.

It is not difficult to show that the metric (\ref{deltametric}) is
regular in the sense of Ref. \cite{gt87}, thus all the curvature
tensor fields make sense as distributions. Taking the
distributional limit as $\delta \to 0$ of
(\ref{deltag11},\ref{deltag00}) we obtain \begin{equation}
 \lim_{\delta\to 0}
G^\xi_{\,\,\xi} = 6 \alpha^2 ;\quad  \lim_{\delta\to 0} G^t_{\,\,t}
= 6 \alpha^2 - 3\alpha \frac{(2s -1)}{s}
 \frac{\left[\Gamma(1 - \frac{1}{2s})\right]^2}{ \Gamma(2 -
\frac{1}{s})}  \, \delta(\xi)\\ \label{zeroeinstein} \end{equation}
corresponding to an infinitely thin domain wall located at $\xi=0$
embedded in a AdS$_5$ spacetime. However, for $\delta\to 0$
(\ref{deltametric}) is not a regular metric in the differentiable 
structure arising from the given chart, and we cannot use the
approximation theorems of \cite{gt87} in order to relate the limit
of the curvature tensor distributions with the limit of the metric
tensor field. Whether or not a metric is regular depends in
general on the differentiable structure imposed on the underlying
manifold. A different
chart may exist for which the resulting differentiable structure
gives a regular metric, but this is of no concern to us here.

\section{Summary and Discussion}

Thick domain wall solutions are not uniquely determined by the scalar
field potential and the boundary conditions on the field at spatial
infinity, but depend also on the subsidiary conditions imposed on the
spacetime metric. We have shown that a theory with a given scalar
field potential admits in general two kinds of solutions, depending on
whether or not  one demands reflection symmetry on the wall plane. If
an appropriate coordinate chart is chosen, the scalar field looks the
same in both solutions. However, their spacetimes are intrinsically
different and cannot be related by a global coordinate change. This is
readily seen when comparing curvature scalars for both
cases. Solutions with reflection symmetry have been shown
 to have a time-dependent
metric, while the asymmetric ones are static. Asymmetric solutions
 are asymptotically flat on one side of the
wall, and become the Taub spacetime on the other side. This result is
 valid for D-2 walls in D dimensions.

By appropriately choosing the coordinate chart, we have shown that
the Einstein equations for both cases can be solved by the same
strategy, namely the appropriate  scaling of vacuum solutions,
allowing to associate an asymmetric solution to any dynamic one.
 Using the method of \cite{gmp02} for generating thick wall
solutions by scaling thin wall (vacuum) solutions, we have given examples
of this, and extend results on the thin-wall limit of dynamic
thick wall solutions to the asymmetric case.

A different way of scaling thin wall solutions that also provides
thick solutions has been presented, and shown to provide exact
solutions o the Einstein-scalar field equations for both cases. As an
example, we have found a family of static, symmetric, 
 ``double'' wall solutions, which contains as a particular case a
previously known BPS solution. In the thin-wall limit, the energy density
and pressure of these walls correspond to a single infinitely thin
sheet.

How four-dimensional gravity arises on non-singular domain walls
or thick 3-brane models has been considered in various five-dimensional models
with Z$_2$ symmetry \cite{dfgk00,g99,kks02,w02}. It would be 
interesting to analyze
the metric fluctuations in the Z$_2$-symmetric case of the double
domain wall spacetimes with metrics given by (\ref{deltametric}).
On the other hand, is the spectrum of general linearized tensor
 fluctuations of the  asymmetric walls consistent
with four-dimensional gravity on the wall? We leave this interesting
questions for a future publication.

\subsection*{Acknowledgments}
We  thank H. Rago for enlightening discussions.  This work
was supported by CDCHT-ULA (Project C-1073-01-05-A) and FONACIT
(Project S1-2000000820). A.M. and A.S. wish to thank ICTP for
hospitality during the completion of this work.

%\bibliographystyle{../biblio/utphys.bst}

%\bibliography{../biblio/all}

\providecommand{\href}[2]{#2}\begingroup\raggedright\endgroup

\end{document}